\documentclass[aps,prl,twocolumn,superscriptaddress]{revtex4-1}
\usepackage{graphicx}
\usepackage{color}
\usepackage{amsfonts}
\usepackage{amsmath}
\usepackage{comment}
\hypersetup{
colorlinks=true,
linkcolor=blue,
citecolor=blue,
urlcolor=blue
}

\begin{document}

\title{Atto-Nm Torque Sensing with a Macroscopic Optomechanical Torsion Pendulum}

\author{Kentaro Komori}
\email[]{komori@granite.phys.s.u-tokyo.ac.jp}
\affiliation{Department of Physics, University of Tokyo, Bunkyo, Tokyo 113-0033, Japan}
\affiliation{LIGO Laboratory, Massachusetts Institute of Technology, Cambridge, MA 02139, USA}
\author{Yutaro Enomoto}
\affiliation{Department of Physics, University of Tokyo, Bunkyo, Tokyo 113-0033, Japan}
\author{Ching Pin Ooi}
\affiliation{Department of Physics, University of Tokyo, Bunkyo, Tokyo 113-0033, Japan}
\author{Yuki Miyazaki}
\affiliation{Department of Physics, University of Tokyo, Bunkyo, Tokyo 113-0033, Japan}
\author{Nobuyuki Matsumoto}
\affiliation{Frontier Research Institute for Interdisciplinary Sciences, Tohoku University, Sendai 980-8578, Japan}
\affiliation{Research Institute of Electrical Communication, Tohoku University, Sendai 980-8577, Japan}
\affiliation{JST, PRESTO, Kawaguchi, Saitama 332-0012, Japan}
\author{Vivishek Sudhir}
\affiliation{LIGO Laboratory, Massachusetts Institute of Technology, Cambridge, MA 02139, USA}
\author{Yuta Michimura}
\affiliation{Department of Physics, University of Tokyo, Bunkyo, Tokyo 113-0033, Japan}
\author{Masaki Ando}
\affiliation{Department of Physics, University of Tokyo, Bunkyo, Tokyo 113-0033, Japan}

\date{\today}

\begin{abstract}
Precise measurements of the displacement of, and force acting on, a mechanical oscillator can be performed by coupling the oscillator to an optical cavity. Brownian thermal forces represent a fundamental limit to measurement sensitivity which impedes the ability to use precise force measurements as a tool of fundamental enquiry, particularly in the context of macroscopic quantum measurements and table-top gravitational experiments. A torsion pendulum with a low mechanical resonant frequency can be limited by very small thermal forces -- from its suspensions -- at frequencies above resonance. Here, we report torque sensing of a 10-mg torsion pendulum formed by a bar mirror, using two optical cavities on either edge. The rotational mode was measured by subtracting the two signals from the cavities, while intracavity radiation pressure forces were used to trap the torsional mode with a 1\,kHz optical spring. The resulting torque sensitivity of 20\,aNm/$\sqrt{\mathrm{Hz}}$ is a record for a milligram scale torsion pendulum. This allows us to test spontaneous wave function collapse in a parameter regime that falls in between that tested by space-based experiments, and high-frequency cryogenic cantilevers.
\end{abstract}

\maketitle

\emph{Introduction.--} The lightly damped torsion pendulum is one of the most sensitive mechanical force detectors~\cite{Gillies1993, Adelberger2009}. In various guises, it has been used throughout the history of precision experimental physics: Cavendish's measurement of the Newtonian gravitational constant using a torsion pendulum~\cite{Cavendish1798} is essentially still the preferred method~\cite{Gillies1997}; the first measurements of the feeble radiation pressure force~\cite{Lebedew1901} and torque~\cite{Beth1936} used one; table-top tests of the equivalence principle~\cite{Dicke1961, Wagner2012}, as well as tests of the inverse-square law of gravity~\cite{Adelberger2003}, continue to employ torsion pendula read-out using optical levers.

In recent years, the ability to enhance the sensitivity of the read-out by integrating mechanical oscillators with optical cavities -- within the field of cavity optomechanics~\cite{Aspelmeyer2014} -- has renewed interest in fundamental tests of quantum mechanics~\cite{Stickler2018}, decoherence mechanisms~\cite{Ghirardi1986, Penrose1996, Marshall2003, Diosi2015, Bassi2017}, and gravitational physics~\cite{Pikovski2012, Kafri2015, Schmole2016, Helou2017, Balushi2018} using table-top mechanical experiments. In general, optomechanical experiments to test several of these proposals call for low frequency (sub-Hz), high quality factor, massive (sub-gram) mechanical oscillators measured with high sensitivity~\cite{Nimmrichter2014}. In this context, a macroscopic torsion pendulum -- a well-proven platform for high-precision mechanical experiments -- coupled to an optical cavity -- for enhanced optical read-out and control -- affords a unique combination of capabilities that could enable quantum state preparation in the future.

In addition to read-out sensitivity, all measurements of the angular displacement of a torsion pendulum are limited by thermal torque fluctuations. For a torsion pendulum at temperature $T$, with a moment of inertia $I$, and damping rate $\gamma_{\mathrm m} (\omega)$, the thermal torque at frequency $\omega$ is quantified by the spectral density~\cite{McCombie1953, Gonzalez1995},
\begin{equation}
\label{eq:thermaltorq}
S_\tau^{\mathrm {th}}(\omega) = 4 k_B T I \gamma_{\mathrm m} (\omega)
\end{equation}
Once extraneous technical noise sources are eliminated, the sensitivity of a torsion pendulum can be limited by thermal noise. One route to further improvement is to employ nano-scale (low moment of inertia) torsional oscillators~\cite{Kuhn2017, Ahn2018}, with low dissipation (low damping rate), operated at cryogenic temperatures~\cite{Kim2016}. Typical nano-scale oscillators operating at high resonant frequencies ($\omega_{\mathrm m}$) tend to be viscously damped, i.e. $\gamma_{\mathrm m}(\omega) = \omega_{\mathrm m}/Q_{\mathrm m}$ is frequency-independent and characterized by the quality factor $Q_{\mathrm m}$. A pico-gram-scale torsional oscillator with high-$Q_{\mathrm m}$, coupled to a cryogenic optical micro-cavity, has realized a record torque sensitivity of $10^{-24}\, \mathrm{Nm/\sqrt{Hz}}$~\cite{Kim2016}.

\begin{figure*}
\centering
\includegraphics[width=\hsize]{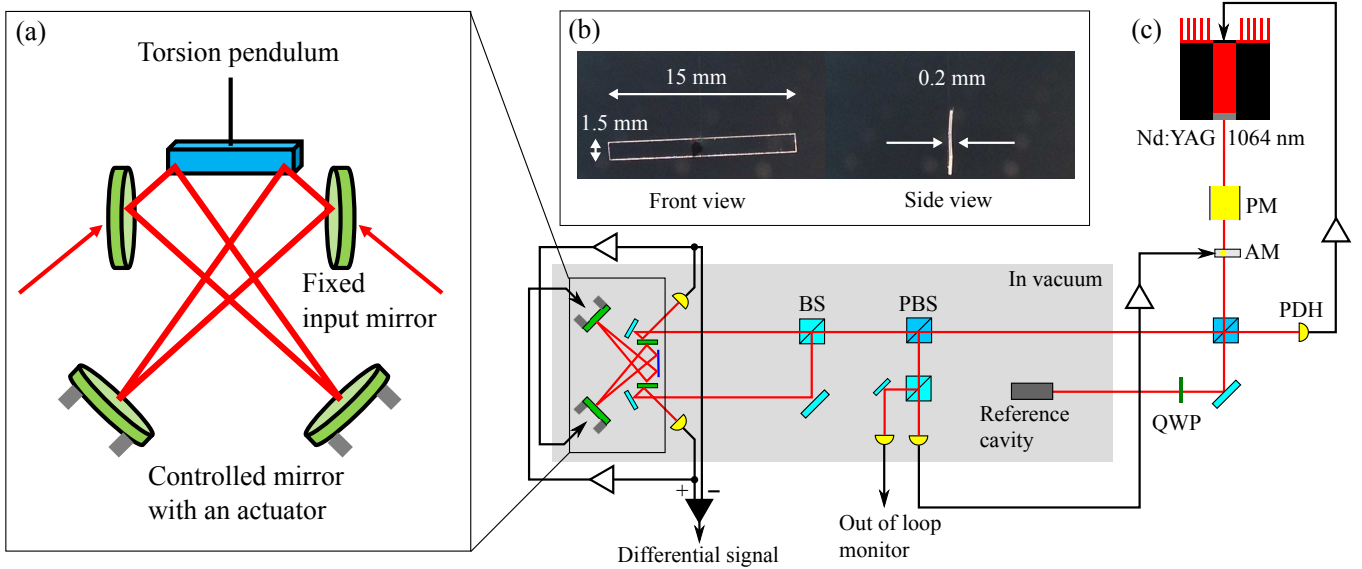}
\caption{
(a) Configuration of the torsion pendulum with two triangular cavities, one on each edge of the bar, used for readout. Input mirrors for either cavity are rigidly mounted, while the third mirror is suspended and can be actuated for cavity length control. (b) Pictures of the bar mirror from front and side views. (c) Simplified schematic of the experiment. Laser frequency is stabilized by a reference cavity with a phase modulator (PM), a polarizing beam splitter (PBS) and a quarter wave plate (QWP) using a Pound-Drever-Hall (PDH) error signal. Laser intensity stabilization is done by feedback to an amplitude modulator (AM). The reflection from the cavities are monitored by photo diodes (PDs) to control the cavity length. The subtracted error signal is analyzed to estimate torsional motion of the bar.
}
\label{figure1}
\end{figure*}

Sub-gram-scale torsional oscillators -- such as the ones required for fundamental tests of decoherence~\cite{Bassi2017, Nimmrichter2014} -- have either been traditionally read-out using an optical lever~\cite{Haiberger2007, Mueller2008b, Cavalleri2009}, or when integrated with an optical cavity, limited by thermal and technical noises~\cite{Mueller2008a, McManus2017}. To the best of our knowledge, the state-of-the-art is a torque sensitivity of $4 \times 10^{-16}\, \mathrm{Nm/\sqrt{Hz}}$ at 6\,kHz (inferred from the reported torque variance of $2 \times 10^{-18}\, \mathrm{Nm}$ achieved after an integration time of 14 hours)~\cite{Haiberger2007}. In this Rapid Communication, we report an order of magnitude improvement in sensitivity to $2 \times 10^{-17}\, \mathrm{Nm/\sqrt{Hz}}$ at around 100\,Hz, with a milligram-scale torsion pendulum at room temperature. Our oscillator is a 10-mg suspended bar-shaped mirror oscillating at 90\,mHz. Two optical cavities are constructed, one on each edge of the bar mirror, in order to sense the rotational mode by subtracting the two signals. Radiation pressure forces from the two cavity modes are used to stiffen the torsional mode using an optical spring, which further dilutes the contribution of suspension thermal noise~\cite{Tan2015}. The low suspension thermal noise has enabled the realization of the best torque sensitivity at these mass scales to our knowledge. This is a crucial step towards tests of gravitational and quantum physics with torsion pendula.

\emph{Concept and experiment.--} At the low frequencies of the suspensions of large oscillators, structural damping leads to a frequency-dependent dissipation~\cite{Gonzalez1995, Li2018}, i.e. $\gamma_{\mathrm m}(\omega)=(\omega_{\mathrm m}/Q_{\mathrm m})(\omega_{\mathrm m}/\omega)$. Thus, for large-scale torsion pendula, thermal torque fluctuations,
\begin{equation}
\label{eq:thermaltorqst}
S_\tau^\mathrm{th,struct}(\omega) = 4k_{\mathrm B} T I \frac{\omega_{\mathrm m}^2}{Q_{\mathrm m} \omega},
\end{equation}
decrease quadratically with the resonant frequency. Thus, a low-frequency torsion pendulum can efficiently offset the benefits of cryogenic operation (as long as the dissipation in the suspension is not increased). Since the restoring torque of a single suspension wire is proportional to the radius of the wire to the fourth power, the resonant frequency of a torsion pendulum can be dramatically decreased by using ultra-thin suspensions. Furthermore, thermal noise is reduced by a $1/\omega$ factor above resonance compared to a viscously damped oscillator.

The torsion pendulum we use, shown in Fig.~\ref{figure1}b, is formed by suspending a 10 mg bar mirror (silica substrate of dimensions 15 mm $\times$ 1.5 mm $\times$ 0.2 mm) on a single strand of carbon fiber that is $\sim 6\,\mu$m thick and 2.2 cm long. The small diameter and low shear modulus of the fiber ($\sim 10$\,GPa) gives a torsion resonant frequency $\omega_{\mathrm m}=2\pi \times 90$\,mHz, while ring-down measurements indicate a damping rate of $\gamma_{\mathrm m}(\omega_{\mathrm m})=2\pi \times (35 \pm 3)\,\mu$Hz, corresponding to a quality factor of $Q_{\mathrm m} =(2.6 \pm 0.2) \times 10^3$. Taken together, Eq.~(\ref{eq:thermaltorqst}) predicts a thermal torque of $0.8\times 10^{-18}$\,Nm/$\sqrt{\mathrm{Hz}}$ at 100\,Hz.

In order to access this state-of-the-art sensitivity at the milligram scale, we use two optical cavities situated on either end of the bar mirror (see Fig.~\ref{figure1}a) to sense its displacement. The cavities have a triangular configuration to quell radiation pressure torque instabilities \cite{Matsumoto2014}. Their input mirrors, half-inch in size with reflectivity $R_\mathrm{i}=99.8\%$, are mounted on a rigid holder which are themselves on picomotors so as to allow for cavity alignment. The second mirror is of a similar size, but with a reflectivity of $R_\mathrm{c}=99.99\%$, is embedded in a brass spacer weighing 70\,g and suspended from four piano wires. It is actuated using a coil-magnet arrangement for cavity length control. The cavities are each driven by 20\,mW of 1064\,nm light from a Nd:YAG laser derived at a beam-splitter (see Fig.~\ref{figure1}c). Either cavity is designed to be 9\,cm long, and their finesse was measured to be $\mathcal{F}_{\mathrm A} = (3.0 \pm 0.3)\times 10^3$ and $\mathcal{F}_{\mathrm B} = (2.4 \pm 0.2) \times 10^3$ respectively. The resulting circulating power $P_\mathrm{A,B} \sim 10$\,W, together with the effective length of the bar $L_{\mathrm{eff}}=10$\,mm (which quantifies the effect of spot position on the bar mirror) allows in principle, a torque sensitivity of $1.2\times 10^{-18}\,\mathrm{Nm}/\sqrt{\mathrm{Hz}}$ at 100\,Hz.

However, various sources of extraneous technical noises need to be mitigated to realize the design potential. Isolation from ground motion is achieved by placing the cavities on an aluminum plate on a double pendulum system and elastomer dampers (Viton) at the bottom; this provides 70\,dB isolation from vertical ground motion and an estimated 100\,dB in the horizontal direction at 100\,Hz. Extraneous laser frequency noise is suppressed by stabilizing the laser with an in-vacuum reference cavity (4.4 cm long with a finesse of $6.4\times 10^4$) using a Pound-Drever-Hall error signal that is used to actuate  the laser piezo; this results in a residual frequency noise of 0.03\,Hz/$\sqrt{\mathrm{Hz}}$ at 100\,Hz, as estimated from the suppressed error signal of the frequency stabilization loop. Laser intensity is actively stabilized to a relative shot noise level of 3 using an amplitude modulator to actuate on a photo detector (PD) signal that monitors laser power. Both the reference cavity and intensity monitor PDs are seismically isolated. The entire experiment, together with the extraneous noise sensors, is operated at a pressure of $2.4\times 10^{-4}$\,Pa to eliminate coupling of acoustic noise, and noise due to residual gas.

To measure the torsional mode, each of the two cavities is locked to the laser. Laser-cavity detuning is controlled by adding offsets to the feedback signal. A radiation pressure optical spring, implemented by blue-detuning the laser to $1/\sqrt{3}\sim0.6$ of the cavity linewidth (to maximize the optical spring for given input power), further suppresses the influence of low-frequency ground vibrations. Finally, amplitude fluctuations of light reflected from each cavity, detected on a PD, are combined to obtain the rotational motion of the torsion pendulum. Common-mode rejection from combining the displacements of the bar's ends further suppresses seismic noise in the signal.

\begin{figure}
\centering
\includegraphics[width=\hsize]{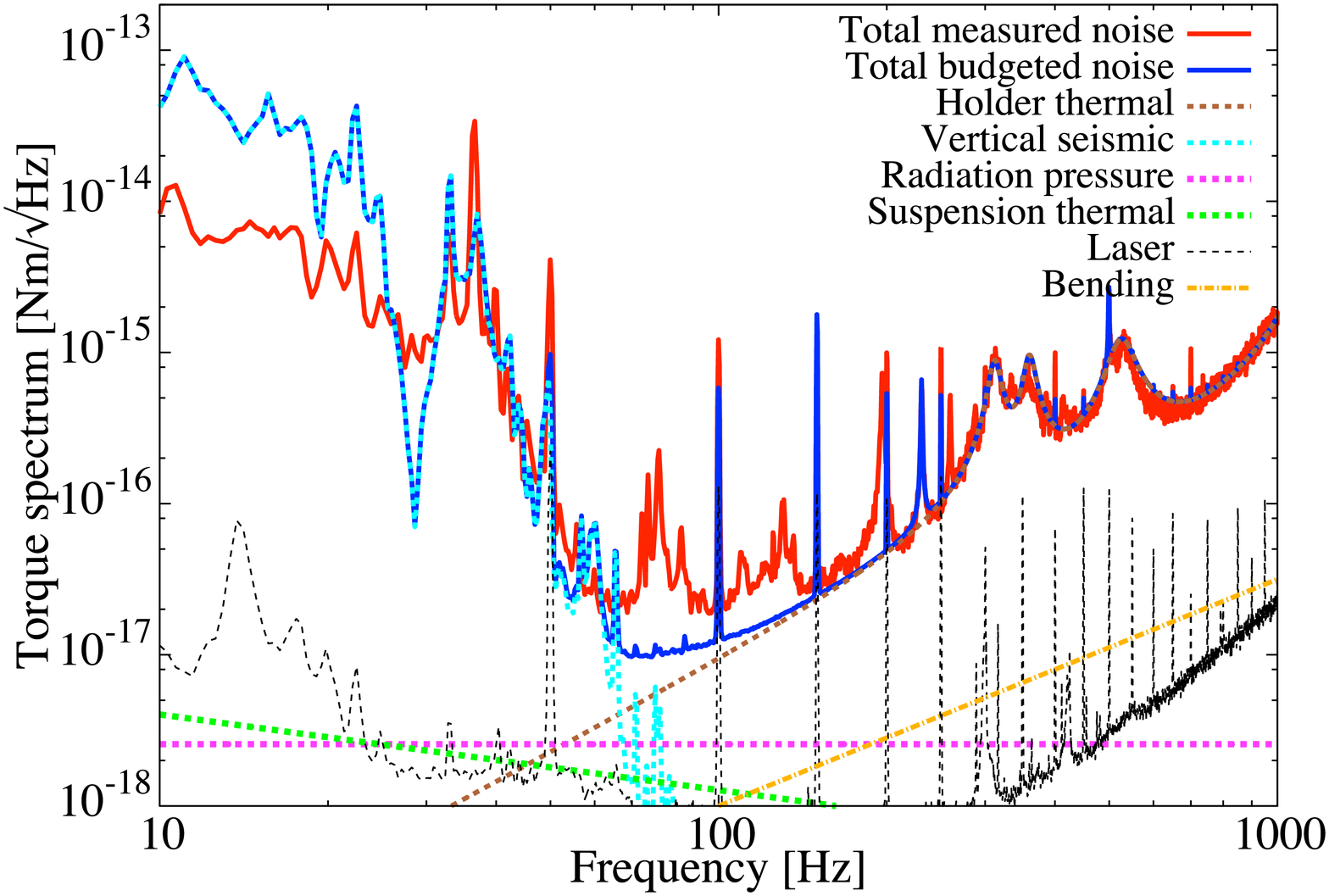}
\caption{
Noise spectra of differential force on the ends of the bar mirror, calibrated as a torque. The measured spectrum (red) can be largely understood using a noise budget (blue) consisting of vertical seismic noise (cyan dotted), thermal noise of the input mirror holder's structural modes (brown dotted), laser noise (black dotted), internal bending mode of the bar mirror (orange dotted), thermal noise from the pendulum suspension (green dotted), and quantum radiation pressure torque noise from the readout cavity modes (red dotted).
}
\label{figure2}
\end{figure}

\emph{Result and discussion.--} The measured torque spectral density is shown in Fig.~\ref{figure2} (red trace). The spectrum is calibrated in torque units using the known susceptibility of the torsion pendulum and the estimated spectrum of the pendulum's angular displacement. The latter is obtained by combining either cavity's displacement noise in a linear superposition, $\theta = (x_\mathrm{B}-\alpha x_\mathrm{A})/L_\mathrm{eff}$. Here, $L_\mathrm{eff} = 10$\,mm is the effective length of the bar inferred from monitoring the bar's bending modes. The superposition coefficient $\alpha$ is ideally unity; however differences in the two beam spot sizes, reflectivities of the two ends of the bar, and transduction of either cavity implies that $\alpha \neq 1$. In order to determine $\alpha$, we use the fact that common-mode extraneous noises, such as external vibration noise, should not couple to the angular displacement of the bar. We are thus led to choose $\alpha = 0.88$ which minimizes the transduction of the vibration noise peak at 73\,Hz (originating from a vacuum pump) by a factor of 0.03. This optimal choice also suppresses other broadband noises in the 50-100\,Hz interval by a factor of 0.5, resulting in a torque sensitivity of $2\times 10^{-17}$\,Nm/$\sqrt{\mathrm{Hz}}$ (corresponding to an angular displacement of $10^{-15}\,\mathrm{rad}/\sqrt{\mathrm{Hz}}$). This is by approximately an order of magnitude above the ideal thermal noise limit expected for the given configuration, and the best torque sensitivity achieved at the milligram scale to our knowledge.

The blue trace in Fig.~\ref{figure2} shows the total budgeted noise, while the various dotted lines show the dominant components of the budget. Vertical seismic noise (cyan dotted), measured independently by a seismometer, limits the sensitivity at low frequencies ($< 50$\,Hz) where the seismometer and the cavity length signals have a coherence close to unity. The transduction from vertical ground motion to cavity length is frequency-dependent, via the 20\,Hz resonance of the vibration isolation platform; in the noise budget, we model the transduction above 20\,Hz. Thermal noise of the input mirror holder (brown dotted) limits sensitivity at high frequencies ($> 150$\,Hz). The non-monolithic construction of the input mirror holder, partly due to the various moving parts for cavity alignment, has several structural resonances which we model using their quality factors (typically below 10) and resonant frequencies. The origin of the mirror holder noise has also been verified by using a single linear cavity, eliminating the rigid input mirror holder. A more fundamental, but small, contribution to the noise budget at these frequencies is due to the thermal motion of the bending mode of the bar mirror substrate (orange dotted)~\cite{thermalnoisecal}. Note that other thermal noises are negligible; this includes higher order bending modes, linear displacements of the suspension, and violin modes of the fiber. Finally, torque-equivalent laser noise (black dotted), arising from residual amplitude noise that limits readout sensitivity (in cavity reflection) and excess frequency noise that gives rise to intra-cavity amplitude noise (for detuned operation) and thus classical radiation pressure torque fluctuations, remain yet smaller. Despite accounting for all the above sources of noise, an unexplained excess remains in the 50-150\,Hz interval; we conjecture that the excess is due to stray scattering of light in the readout path.

\begin{figure}
\centering
\includegraphics[width=\hsize]{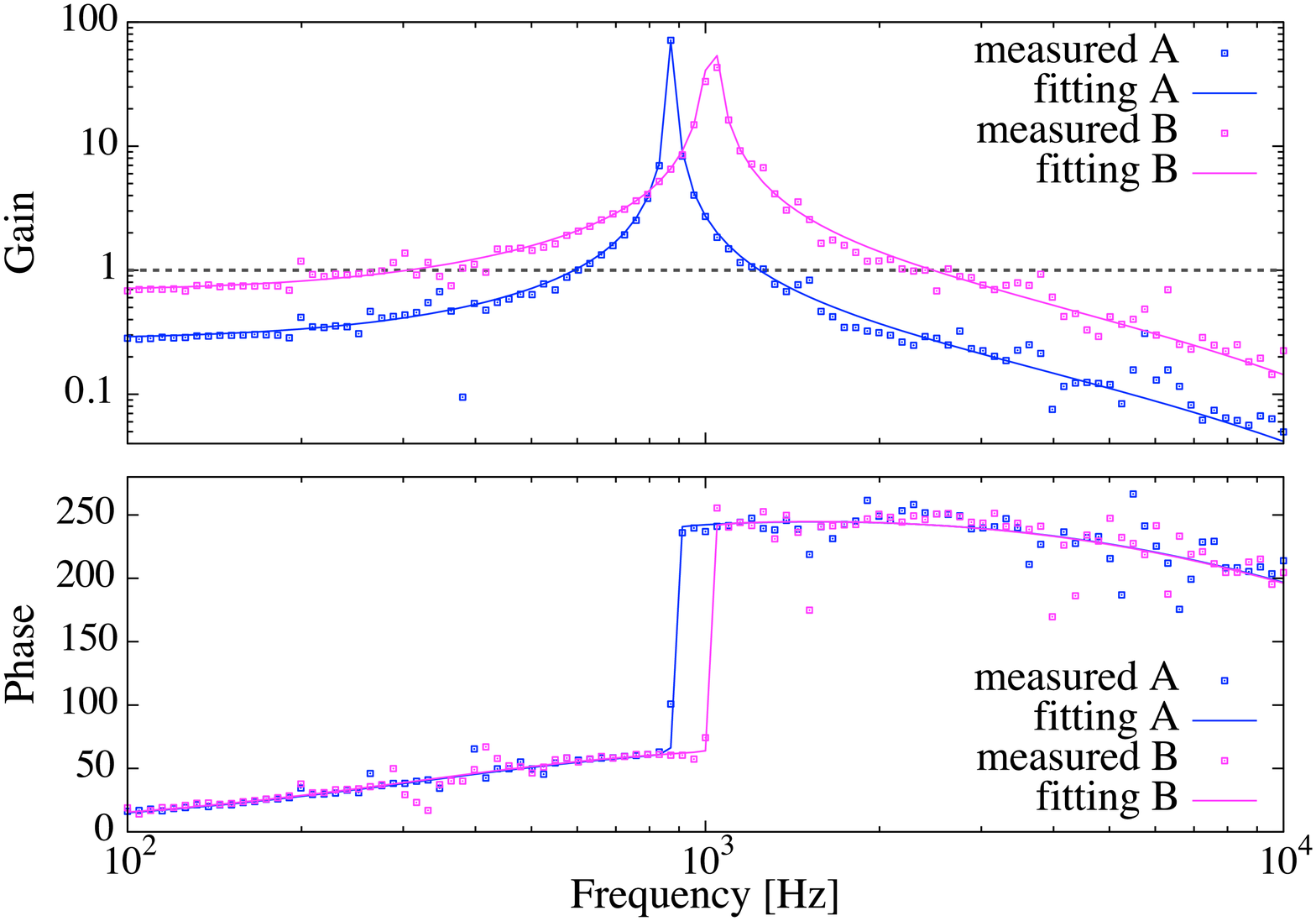}
\caption{
Optical spring response of either cavity, measured as the open loop transfer function of the respective cavity's length control signal. The measured amplitude and phase of the cavity A (B) are shown as blue (purple) dots, while solid lines show a model fitted to the data.
}
\label{figure3}
\end{figure}

Fundamental noises related to the pendulum, such as suspension thermal noise of the torsional mode, and quantum radiation pressure noise, are shown as the green and red dotted lines respectively. They are plotted as torques acting on the measured beam spot positions. Suspension thermal noise, calculated from Eq.~(\ref{eq:thermaltorqst}) using the measured resonant frequency and quality factor, is below the total budgeted by an order of magnitude at 100\,Hz. Quantum radiation pressure torque noise is given by
\begin{equation}
\label{eq:radtorq}
S^{\mathrm{rad}}_{\tau}(\omega) = \frac{2\hbar}{\delta} m \omega_{\mathrm{eff}}^2 L_{\mathrm{eff}}^2,
\end{equation}
where $\delta$ is the cavity detuning normalized by the cavity line width, and $\omega_{\mathrm{eff}}$ is the effective mechanical resonant frequency including the optical spring. In our experiment, $\delta \sim 0.6$, while the effective mechanical frequency, dominated by the optical spring (see Fig.~\ref{figure3}), is $\omega_\mathrm{eff} \sim 2\pi \times 1$\,kHz. The resulting estimate for the quantum radiation pressure torque noise contributes about $(14\pm 3)$\% to the measured spectrum around 100\,Hz. This value, along with~\cite{Matsumoto2019}, is the highest reported on milligram and gram-scale experiments to observe quantum radiation pressure noise~\cite{Neben2012, Matsumoto2015, Nguyen2015}.

\begin{figure}
\centering
\includegraphics[width=\hsize]{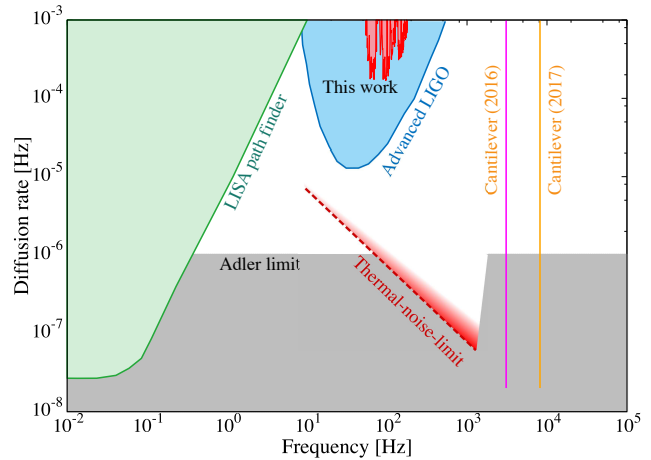}
\caption{
Landscape of CSL diffusion rate excluded by mechanical experiments. The gray region represents the range of diffusion parameters yet to be tested~\cite{Adler2007}. Resonant thermal noise measurements using cryogenic cantilevers~\cite{Vinante2016, Vinante2017} exclude a narrow range of parameters at 1-10\,kHz. Advanced LIGO's sensitivity does not yet broach the interesting region~\cite{Carlesso2016}, while LISA Path Finder does exclude swaths at low frequency~\cite{Helou2017b}. Red region at the top represents the current experiment, while the red dashed line is the projection for the current experiment if it were limited by the suspension thermal noise.
}
\label{figure4}
\end{figure}

The achieved torque sensitivity can be cast as a test of spontaneous wave function collapse~\cite{Bassi2017, Diosi2015}. The CSL model for wave function collapse posits an additional torque noise on the oscillator~\cite{Diosi2015},
\begin{equation}
\label{eq:csltorq}
S^{\mathrm{CSL}}_{\tau}(\omega) =  \lambda_{\mathrm{CSL}} \frac{8\pi \hbar^2 r_{\mathrm{CSL}}^2}{m_0^2} \frac{\rho I}{d},
\end{equation}
where $\lambda_{\mathrm{CSL}}$ and $r_{\mathrm{CSL}}$ are the CSL parameters quantifying the excess momentum diffusion rate and characteristic length; $m_0 \simeq 1.66\times 10^{-27}$\,kg is the atomic mass unit, $\rho$ the density, and $d$ the thickness of the torsional oscillator. The torque sensitivity achieved in our experiment can therefore be interpreted as setting a bound on the diffusion rate $\lambda_{\mathrm{CSL}}$. The red trace in Fig.~\ref{figure4} shows the limit set by the current experiment. Interestingly, our experiment straddles the region of parameter space left untested by LISA Path Finder~\cite{Helou2017b}, and high-frequency cantilevers~\cite{Vinante2017}. In fact, with further improvement to reach the thermal-noise limit of our torsion pendulum, we can bridge this gap (red dashed line), and exclude some of the region yet to be tested (gray). The low mass and low thermal noise allow us to improve upon the limit set by advanced LIGO~\cite{Carlesso2016} in this region of parameter space.

\emph{Future directions and conclusion.--} The current experiment is limited by thermal noise from the low order bending modes of the bar, and radiation-pressure-induced cavity misalignment at higher optical powers. The former problem can be solved by replacing the bar mirror with a dumbbell-shaped arrangement of mirrors at the two ends. The shorter and thicker substrate in this case increases the resonant frequencies of the bending modes and reduces their thermal noise. In ongoing work, we are experimenting with a linear cavity - without the fixed mirror - to readout the oscillator's displacement in a manner that is resilient to radiation pressure torque instabilities; this would eliminate thermal noise due to the mirror holder. In this fashion, extraneous noise above 150\,Hz is expected to be mitigated, and lead to a torque sensitivity at the standard quantum limit, about $1\times 10^{-18}$\,Nm/$\sqrt{\mathrm {Hz}}$ at 400\,Hz.

In conclusion, we have demonstrated a milligram-scale torsion pendulum with the best torque sensitivity at these mass scales. This is achieved by employing a suspension made of a single strand of carbon fiber featuring an exceptionally small resonant frequency with internal dissipation that is structural in nature. By forming triangular cavities on each edge of the torsion bar -- to eliminate radiation pressure torque instabilities -- and performing differential readout of the output of the two cavities, we realize a torque sensitivity at the few atto-Nm level. This is within an order of magnitude of being limited by quantum radiation pressure torque fluctuations, and has the potential to explore various aspects of gravitational and quantum physics, including stringent tests of spontaneous wave function collapse.

\emph{Acknowledgements.--} We would like to acknowledge T. Shimoda, K. Nagano, Y. Yoneta, and M. Evans for fruitful discussions. We also thank OptoSigma company for producing the bar mirror, TORAY company for making the carbon fiber, and S. Otsuka and T. Shimozawa for manufacturing the mechanical parts. This work was supported by JSPS KAKENHI Grant No. 16J01010, JST CREST Grant No. JPMJCR1873, and MEXT Quantum LEAP Flagship Program (MEXT Q-LEAP) Grant Number JPMXS0118070351. VS is supported by the Swiss National Science Foundation fellowship grant P2ELP2\_178231. This paper has LIGO Document Number LIGO-P1900223.

\bibliography{paper}

\begin{thebibliography}{45}%
\makeatletter
\providecommand \@ifxundefined [1]{%
 \@ifx{#1\undefined}
}%
\providecommand \@ifnum [1]{%
 \ifnum #1\expandafter \@firstoftwo
 \else \expandafter \@secondoftwo
 \fi
}%
\providecommand \@ifx [1]{%
 \ifx #1\expandafter \@firstoftwo
 \else \expandafter \@secondoftwo
 \fi
}%
\providecommand \natexlab [1]{#1}%
\providecommand \enquote  [1]{``#1''}%
\providecommand \bibnamefont  [1]{#1}%
\providecommand \bibfnamefont [1]{#1}%
\providecommand \citenamefont [1]{#1}%
\providecommand \href@noop [0]{\@secondoftwo}%
\providecommand \href [0]{\begingroup \@sanitize@url \@href}%
\providecommand \@href[1]{\@@startlink{#1}\@@href}%
\providecommand \@@href[1]{\endgroup#1\@@endlink}%
\providecommand \@sanitize@url [0]{\catcode `\\12\catcode `\$12\catcode
  `\&12\catcode `\#12\catcode `\^12\catcode `\_12\catcode `\%12\relax}%
\providecommand \@@startlink[1]{}%
\providecommand \@@endlink[0]{}%
\providecommand \url  [0]{\begingroup\@sanitize@url \@url }%
\providecommand \@url [1]{\endgroup\@href {#1}{\urlprefix }}%
\providecommand \urlprefix  [0]{URL }%
\providecommand \Eprint [0]{\href }%
\providecommand \doibase [0]{https://doi.org/}%
\providecommand \selectlanguage [0]{\@gobble}%
\providecommand \bibinfo  [0]{\@secondoftwo}%
\providecommand \bibfield  [0]{\@secondoftwo}%
\providecommand \translation [1]{[#1]}%
\providecommand \BibitemOpen [0]{}%
\providecommand \bibitemStop [0]{}%
\providecommand \bibitemNoStop [0]{.\EOS\space}%
\providecommand \EOS [0]{\spacefactor3000\relax}%
\providecommand \BibitemShut  [1]{\csname bibitem#1\endcsname}%
\let\auto@bib@innerbib\@empty
\bibitem [{\citenamefont {Gillies}\ and\ \citenamefont
  {Ritter}(1993)}]{Gillies1993}%
  \BibitemOpen
  \bibfield  {author} {\bibinfo {author} {\bibfnamefont {G.~T.}\ \bibnamefont
  {Gillies}}\ and\ \bibinfo {author} {\bibfnamefont {R.~C.}\ \bibnamefont
  {Ritter}},\ }\href {https://doi.org/10.1063/1.1144248} {\bibfield  {journal}
  {\bibinfo  {journal} {Review of Scientific Instruments}\ }\textbf {\bibinfo
  {volume} {64}},\ \bibinfo {pages} {283} (\bibinfo {year} {1993})}\BibitemShut
  {NoStop}%
\bibitem [{\citenamefont {Adelberger}\ \emph {et~al.}(2009)\citenamefont
  {Adelberger}, \citenamefont {Gundlach}, \citenamefont {Heckel}, \citenamefont
  {Hoedl},\ and\ \citenamefont {Schlamminger}}]{Adelberger2009}%
  \BibitemOpen
  \bibfield  {author} {\bibinfo {author} {\bibfnamefont {E.}~\bibnamefont
  {Adelberger}}, \bibinfo {author} {\bibfnamefont {J.}~\bibnamefont
  {Gundlach}}, \bibinfo {author} {\bibfnamefont {B.}~\bibnamefont {Heckel}},
  \bibinfo {author} {\bibfnamefont {S.}~\bibnamefont {Hoedl}},\ and\ \bibinfo
  {author} {\bibfnamefont {S.}~\bibnamefont {Schlamminger}},\ }\href
  {https://doi.org/10.1016/j.ppnp.2008.08.002} {\bibfield  {journal} {\bibinfo
  {journal} {Progress in Particle and Nuclear Physics}\ }\textbf {\bibinfo
  {volume} {62}},\ \bibinfo {pages} {102 } (\bibinfo {year}
  {2009})}\BibitemShut {NoStop}%
\bibitem [{\citenamefont {Cavendish}(1798)}]{Cavendish1798}%
  \BibitemOpen
  \bibfield  {author} {\bibinfo {author} {\bibfnamefont {H.}~\bibnamefont
  {Cavendish}},\ }\href {https://doi.org/10.1098/rstl.1798.0022} {\bibfield
  {journal} {\bibinfo  {journal} {Phil. Trans. R. Soc.}\ }\textbf {\bibinfo
  {volume} {88}},\ \bibinfo {pages} {469} (\bibinfo {year} {1798})}\BibitemShut
  {NoStop}%
\bibitem [{\citenamefont {Gillies}(1997)}]{Gillies1997}%
  \BibitemOpen
  \bibfield  {author} {\bibinfo {author} {\bibfnamefont {G.~T.}\ \bibnamefont
  {Gillies}},\ }\href {https://doi.org/10.1088/0034-4885/60/2/001} {\bibfield
  {journal} {\bibinfo  {journal} {Reports on Progress in Physics}\ }\textbf
  {\bibinfo {volume} {60}},\ \bibinfo {pages} {151} (\bibinfo {year}
  {1997})}\BibitemShut {NoStop}%
\bibitem [{\citenamefont {Lebedew}(1901)}]{Lebedew1901}%
  \BibitemOpen
  \bibfield  {author} {\bibinfo {author} {\bibfnamefont {P.}~\bibnamefont
  {Lebedew}},\ }\href {https://doi.org/10.1002/andp.19013111102} {\bibfield
  {journal} {\bibinfo  {journal} {Annalen der Physik}\ }\textbf {\bibinfo
  {volume} {311}},\ \bibinfo {pages} {433} (\bibinfo {year}
  {1901})}\BibitemShut {NoStop}%
\bibitem [{\citenamefont {Beth}(1936)}]{Beth1936}%
  \BibitemOpen
  \bibfield  {author} {\bibinfo {author} {\bibfnamefont {R.~A.}\ \bibnamefont
  {Beth}},\ }\href {https://doi.org/10.1103/PhysRev.50.115} {\bibfield
  {journal} {\bibinfo  {journal} {Phys. Rev.}\ }\textbf {\bibinfo {volume}
  {50}},\ \bibinfo {pages} {115} (\bibinfo {year} {1936})}\BibitemShut
  {NoStop}%
\bibitem [{\citenamefont {Dicke}(1961)}]{Dicke1961}%
  \BibitemOpen
  \bibfield  {author} {\bibinfo {author} {\bibfnamefont {R.~H.}\ \bibnamefont
  {Dicke}},\ }\href {http://www.jstor.org/stable/24937165} {\bibfield
  {journal} {\bibinfo  {journal} {Scientific American}\ }\textbf {\bibinfo
  {volume} {205}},\ \bibinfo {pages} {84} (\bibinfo {year} {1961})}\BibitemShut
  {NoStop}%
\bibitem [{\citenamefont {Wagner}\ \emph {et~al.}(2012)\citenamefont {Wagner},
  \citenamefont {Schlamminger}, \citenamefont {Gundlach},\ and\ \citenamefont
  {Adelberger}}]{Wagner2012}%
  \BibitemOpen
  \bibfield  {author} {\bibinfo {author} {\bibfnamefont {T.~A.}\ \bibnamefont
  {Wagner}}, \bibinfo {author} {\bibfnamefont {S.}~\bibnamefont
  {Schlamminger}}, \bibinfo {author} {\bibfnamefont {J.~H.}\ \bibnamefont
  {Gundlach}},\ and\ \bibinfo {author} {\bibfnamefont {E.~G.}\ \bibnamefont
  {Adelberger}},\ }\href {https://doi.org/10.1088/0264-9381/29/18/184002}
  {\bibfield  {journal} {\bibinfo  {journal} {Classical and Quantum Gravity}\
  }\textbf {\bibinfo {volume} {29}},\ \bibinfo {pages} {184002} (\bibinfo
  {year} {2012})}\BibitemShut {NoStop}%
\bibitem [{\citenamefont {Adelberger}\ \emph {et~al.}(2003)\citenamefont
  {Adelberger}, \citenamefont {Heckel},\ and\ \citenamefont
  {Nelson}}]{Adelberger2003}%
  \BibitemOpen
  \bibfield  {author} {\bibinfo {author} {\bibfnamefont {E.}~\bibnamefont
  {Adelberger}}, \bibinfo {author} {\bibfnamefont {B.}~\bibnamefont {Heckel}},\
  and\ \bibinfo {author} {\bibfnamefont {A.}~\bibnamefont {Nelson}},\ }\href
  {https://doi.org/10.1146/annurev.nucl.53.041002.110503} {\bibfield  {journal}
  {\bibinfo  {journal} {Annual Review of Nuclear and Particle Science}\
  }\textbf {\bibinfo {volume} {53}},\ \bibinfo {pages} {77} (\bibinfo {year}
  {2003})}\BibitemShut {NoStop}%
\bibitem [{\citenamefont {Aspelmeyer}\ \emph {et~al.}(2014)\citenamefont
  {Aspelmeyer}, \citenamefont {Kippenberg},\ and\ \citenamefont
  {Marquardt}}]{Aspelmeyer2014}%
  \BibitemOpen
  \bibfield  {author} {\bibinfo {author} {\bibfnamefont {M.}~\bibnamefont
  {Aspelmeyer}}, \bibinfo {author} {\bibfnamefont {T.~J.}\ \bibnamefont
  {Kippenberg}},\ and\ \bibinfo {author} {\bibfnamefont {F.}~\bibnamefont
  {Marquardt}},\ }\href {https://doi.org/10.1103/RevModPhys.86.1391} {\bibfield
   {journal} {\bibinfo  {journal} {Rev. Mod. Phys.}\ }\textbf {\bibinfo
  {volume} {86}},\ \bibinfo {pages} {1391} (\bibinfo {year}
  {2014})}\BibitemShut {NoStop}%
\bibitem [{\citenamefont {Stickler}\ \emph {et~al.}(2018)\citenamefont
  {Stickler}, \citenamefont {Papendell}, \citenamefont {Kuhn}, \citenamefont
  {Schrinski}, \citenamefont {Millen}, \citenamefont {Arndt},\ and\
  \citenamefont {Hornberger}}]{Stickler2018}%
  \BibitemOpen
  \bibfield  {author} {\bibinfo {author} {\bibfnamefont {B.~A.}\ \bibnamefont
  {Stickler}}, \bibinfo {author} {\bibfnamefont {B.}~\bibnamefont {Papendell}},
  \bibinfo {author} {\bibfnamefont {S.}~\bibnamefont {Kuhn}}, \bibinfo {author}
  {\bibfnamefont {B.}~\bibnamefont {Schrinski}}, \bibinfo {author}
  {\bibfnamefont {J.}~\bibnamefont {Millen}}, \bibinfo {author} {\bibfnamefont
  {M.}~\bibnamefont {Arndt}},\ and\ \bibinfo {author} {\bibfnamefont
  {K.}~\bibnamefont {Hornberger}},\ }\href
  {https://doi.org/10.1088/1367-2630/aaece4} {\bibfield  {journal} {\bibinfo
  {journal} {New J. Phys.}\ }\textbf {\bibinfo {volume} {20}},\ \bibinfo
  {pages} {122001} (\bibinfo {year} {2018})}\BibitemShut {NoStop}%
\bibitem [{\citenamefont {Ghirardi}\ \emph {et~al.}(1986)\citenamefont
  {Ghirardi}, \citenamefont {Rimini},\ and\ \citenamefont
  {Weber}}]{Ghirardi1986}%
  \BibitemOpen
  \bibfield  {author} {\bibinfo {author} {\bibfnamefont {G.~C.}\ \bibnamefont
  {Ghirardi}}, \bibinfo {author} {\bibfnamefont {A.}~\bibnamefont {Rimini}},\
  and\ \bibinfo {author} {\bibfnamefont {T.}~\bibnamefont {Weber}},\ }\href
  {https://doi.org/10.1103/PhysRevD.34.470} {\bibfield  {journal} {\bibinfo
  {journal} {Phys. Rev. D}\ }\textbf {\bibinfo {volume} {34}},\ \bibinfo
  {pages} {470} (\bibinfo {year} {1986})}\BibitemShut {NoStop}%
\bibitem [{\citenamefont {Penrose}(1996)}]{Penrose1996}%
  \BibitemOpen
  \bibfield  {author} {\bibinfo {author} {\bibfnamefont {R.}~\bibnamefont
  {Penrose}},\ }\href {https://doi.org/10.1007/BF02105068} {\bibfield
  {journal} {\bibinfo  {journal} {Gen. Relat. Gravit.}\ }\textbf {\bibinfo
  {volume} {28}},\ \bibinfo {pages} {581} (\bibinfo {year} {1996})}\BibitemShut
  {NoStop}%
\bibitem [{\citenamefont {Marshall}\ \emph {et~al.}(2003)\citenamefont
  {Marshall}, \citenamefont {Simon}, \citenamefont {Penrose},\ and\
  \citenamefont {Bouwmeester}}]{Marshall2003}%
  \BibitemOpen
  \bibfield  {author} {\bibinfo {author} {\bibfnamefont {W.}~\bibnamefont
  {Marshall}}, \bibinfo {author} {\bibfnamefont {C.}~\bibnamefont {Simon}},
  \bibinfo {author} {\bibfnamefont {R.}~\bibnamefont {Penrose}},\ and\ \bibinfo
  {author} {\bibfnamefont {D.}~\bibnamefont {Bouwmeester}},\ }\href
  {https://doi.org/10.1103/PhysRevLett.91.130401} {\bibfield  {journal}
  {\bibinfo  {journal} {Phys. Rev. Lett.}\ }\textbf {\bibinfo {volume} {91}},\
  \bibinfo {pages} {130401} (\bibinfo {year} {2003})}\BibitemShut {NoStop}%
\bibitem [{\citenamefont {Di\'osi}(2015)}]{Diosi2015}%
  \BibitemOpen
  \bibfield  {author} {\bibinfo {author} {\bibfnamefont {L.}~\bibnamefont
  {Di\'osi}},\ }\href {https://doi.org/10.1103/PhysRevLett.114.050403}
  {\bibfield  {journal} {\bibinfo  {journal} {Phys. Rev. Lett.}\ }\textbf
  {\bibinfo {volume} {114}},\ \bibinfo {pages} {050403} (\bibinfo {year}
  {2015})}\BibitemShut {NoStop}%
\bibitem [{\citenamefont {Bassi}\ \emph {et~al.}(2017)\citenamefont {Bassi},
  \citenamefont {Gro{\ss}ardt},\ and\ \citenamefont {Ulbricht}}]{Bassi2017}%
  \BibitemOpen
  \bibfield  {author} {\bibinfo {author} {\bibfnamefont {A.}~\bibnamefont
  {Bassi}}, \bibinfo {author} {\bibfnamefont {A.}~\bibnamefont
  {Gro{\ss}ardt}},\ and\ \bibinfo {author} {\bibfnamefont {H.}~\bibnamefont
  {Ulbricht}},\ }\href {https://doi.org/10.1088/1361-6382/aa864f} {\bibfield
  {journal} {\bibinfo  {journal} {Classical Quantum Gravity}\ }\textbf
  {\bibinfo {volume} {34}},\ \bibinfo {pages} {193002} (\bibinfo {year}
  {2017})}\BibitemShut {NoStop}%
\bibitem [{\citenamefont {Pikovski}\ \emph {et~al.}(2012)\citenamefont
  {Pikovski}, \citenamefont {Vanner}, \citenamefont {Aspelmeyer}, \citenamefont
  {Kim},\ and\ \citenamefont {Brukner}}]{Pikovski2012}%
  \BibitemOpen
  \bibfield  {author} {\bibinfo {author} {\bibfnamefont {I.}~\bibnamefont
  {Pikovski}}, \bibinfo {author} {\bibfnamefont {M.~R.}\ \bibnamefont
  {Vanner}}, \bibinfo {author} {\bibfnamefont {M.}~\bibnamefont {Aspelmeyer}},
  \bibinfo {author} {\bibfnamefont {M.~S.}\ \bibnamefont {Kim}},\ and\ \bibinfo
  {author} {\bibfnamefont {{\v C}.}~\bibnamefont {Brukner}},\ }\href
  {https://doi.org/10.1038/nphys2262} {\bibfield  {journal} {\bibinfo
  {journal} {Nature Physics}\ }\textbf {\bibinfo {volume} {8}},\ \bibinfo
  {pages} {393} (\bibinfo {year} {2012})}\BibitemShut {NoStop}%
\bibitem [{\citenamefont {Kafri}\ \emph {et~al.}(2015)\citenamefont {Kafri},
  \citenamefont {Milburn},\ and\ \citenamefont {Taylor}}]{Kafri2015}%
  \BibitemOpen
  \bibfield  {author} {\bibinfo {author} {\bibfnamefont {D.}~\bibnamefont
  {Kafri}}, \bibinfo {author} {\bibfnamefont {G.~J.}\ \bibnamefont {Milburn}},\
  and\ \bibinfo {author} {\bibfnamefont {J.~M.}\ \bibnamefont {Taylor}},\
  }\href {https://doi.org/10.1088/1367-2630/17/1/015006} {\bibfield  {journal}
  {\bibinfo  {journal} {New Journal of Physics}\ }\textbf {\bibinfo {volume}
  {17}},\ \bibinfo {pages} {015006} (\bibinfo {year} {2015})}\BibitemShut
  {NoStop}%
\bibitem [{\citenamefont {Schm{\"o}le}\ \emph {et~al.}(2016)\citenamefont
  {Schm{\"o}le}, \citenamefont {Dragosits}, \citenamefont {Hepach},\ and\
  \citenamefont {Aspelmeyer}}]{Schmole2016}%
  \BibitemOpen
  \bibfield  {author} {\bibinfo {author} {\bibfnamefont {J.}~\bibnamefont
  {Schm{\"o}le}}, \bibinfo {author} {\bibfnamefont {M.}~\bibnamefont
  {Dragosits}}, \bibinfo {author} {\bibfnamefont {H.}~\bibnamefont {Hepach}},\
  and\ \bibinfo {author} {\bibfnamefont {M.}~\bibnamefont {Aspelmeyer}},\
  }\href {https://doi.org/10.1088/0264-9381/33/12/125031} {\bibfield  {journal}
  {\bibinfo  {journal} {Classical Quantum Gravity}\ }\textbf {\bibinfo {volume}
  {33}},\ \bibinfo {pages} {125031} (\bibinfo {year} {2016})}\BibitemShut
  {NoStop}%
\bibitem [{\citenamefont {Helou}\ \emph
  {et~al.}(2017{\natexlab{a}})\citenamefont {Helou}, \citenamefont {Luo},
  \citenamefont {Yeh}, \citenamefont {Shao}, \citenamefont {Slagmolen},
  \citenamefont {McClelland},\ and\ \citenamefont {Chen}}]{Helou2017}%
  \BibitemOpen
  \bibfield  {author} {\bibinfo {author} {\bibfnamefont {B.}~\bibnamefont
  {Helou}}, \bibinfo {author} {\bibfnamefont {J.}~\bibnamefont {Luo}}, \bibinfo
  {author} {\bibfnamefont {H.-C.}\ \bibnamefont {Yeh}}, \bibinfo {author}
  {\bibfnamefont {C.-g.}\ \bibnamefont {Shao}}, \bibinfo {author}
  {\bibfnamefont {B.~J.~J.}\ \bibnamefont {Slagmolen}}, \bibinfo {author}
  {\bibfnamefont {D.~E.}\ \bibnamefont {McClelland}},\ and\ \bibinfo {author}
  {\bibfnamefont {Y.}~\bibnamefont {Chen}},\ }\href
  {https://doi.org/10.1103/PhysRevD.96.044008} {\bibfield  {journal} {\bibinfo
  {journal} {Phys. Rev. D}\ }\textbf {\bibinfo {volume} {96}},\ \bibinfo
  {pages} {044008} (\bibinfo {year} {2017}{\natexlab{a}})}\BibitemShut
  {NoStop}%
\bibitem [{\citenamefont {Al~Balushi}\ \emph {et~al.}(2018)\citenamefont
  {Al~Balushi}, \citenamefont {Cong},\ and\ \citenamefont
  {Mann}}]{Balushi2018}%
  \BibitemOpen
  \bibfield  {author} {\bibinfo {author} {\bibfnamefont {A.}~\bibnamefont
  {Al~Balushi}}, \bibinfo {author} {\bibfnamefont {W.}~\bibnamefont {Cong}},\
  and\ \bibinfo {author} {\bibfnamefont {R.~B.}\ \bibnamefont {Mann}},\ }\href
  {https://doi.org/10.1103/PhysRevA.98.043811} {\bibfield  {journal} {\bibinfo
  {journal} {Phys. Rev. A}\ }\textbf {\bibinfo {volume} {98}},\ \bibinfo
  {pages} {043811} (\bibinfo {year} {2018})}\BibitemShut {NoStop}%
\bibitem [{\citenamefont {Nimmrichter}\ \emph {et~al.}(2014)\citenamefont
  {Nimmrichter}, \citenamefont {Hornberger},\ and\ \citenamefont
  {Hammerer}}]{Nimmrichter2014}%
  \BibitemOpen
  \bibfield  {author} {\bibinfo {author} {\bibfnamefont {S.}~\bibnamefont
  {Nimmrichter}}, \bibinfo {author} {\bibfnamefont {K.}~\bibnamefont
  {Hornberger}},\ and\ \bibinfo {author} {\bibfnamefont {K.}~\bibnamefont
  {Hammerer}},\ }\href {https://doi.org/10.1103/PhysRevLett.113.020405}
  {\bibfield  {journal} {\bibinfo  {journal} {Phys. Rev. Lett.}\ }\textbf
  {\bibinfo {volume} {113}},\ \bibinfo {pages} {020405} (\bibinfo {year}
  {2014})}\BibitemShut {NoStop}%
\bibitem [{\citenamefont {McCombie}(1953)}]{McCombie1953}%
  \BibitemOpen
  \bibfield  {author} {\bibinfo {author} {\bibfnamefont {C.~W.}\ \bibnamefont
  {McCombie}},\ }\href {https://doi.org/10.1088/0034-4885/16/1/307} {\bibfield
  {journal} {\bibinfo  {journal} {Reports on Progress in Physics}\ }\textbf
  {\bibinfo {volume} {16}},\ \bibinfo {pages} {266} (\bibinfo {year}
  {1953})}\BibitemShut {NoStop}%
\bibitem [{\citenamefont {Gonz{\'a}lez}\ and\ \citenamefont
  {Saulson}(1995)}]{Gonzalez1995}%
  \BibitemOpen
  \bibfield  {author} {\bibinfo {author} {\bibfnamefont {G.~I.}\ \bibnamefont
  {Gonz{\'a}lez}}\ and\ \bibinfo {author} {\bibfnamefont {P.~R.}\ \bibnamefont
  {Saulson}},\ }\href
  {https://doi.org/https://doi.org/10.1016/0375-9601(95)00194-8} {\bibfield
  {journal} {\bibinfo  {journal} {Phys. Lett. A}\ }\textbf {\bibinfo {volume}
  {201}},\ \bibinfo {pages} {12 } (\bibinfo {year} {1995})}\BibitemShut
  {NoStop}%
\bibitem [{\citenamefont {Kuhn}\ \emph {et~al.}(2017)\citenamefont {Kuhn},
  \citenamefont {Kosloff}, \citenamefont {Stickler}, \citenamefont {Patolsky},
  \citenamefont {Hornberger}, \citenamefont {Arndt},\ and\ \citenamefont
  {Millen}}]{Kuhn2017}%
  \BibitemOpen
  \bibfield  {author} {\bibinfo {author} {\bibfnamefont {S.}~\bibnamefont
  {Kuhn}}, \bibinfo {author} {\bibfnamefont {A.}~\bibnamefont {Kosloff}},
  \bibinfo {author} {\bibfnamefont {B.~A.}\ \bibnamefont {Stickler}}, \bibinfo
  {author} {\bibfnamefont {F.}~\bibnamefont {Patolsky}}, \bibinfo {author}
  {\bibfnamefont {K.}~\bibnamefont {Hornberger}}, \bibinfo {author}
  {\bibfnamefont {M.}~\bibnamefont {Arndt}},\ and\ \bibinfo {author}
  {\bibfnamefont {J.}~\bibnamefont {Millen}},\ }\href
  {https://doi.org/10.1364/OPTICA.4.000356} {\bibfield  {journal} {\bibinfo
  {journal} {Optica}\ }\textbf {\bibinfo {volume} {4}},\ \bibinfo {pages} {356}
  (\bibinfo {year} {2017})}\BibitemShut {NoStop}%
\bibitem [{\citenamefont {Ahn}\ \emph {et~al.}(2018)\citenamefont {Ahn},
  \citenamefont {Xu}, \citenamefont {Bang}, \citenamefont {Deng}, \citenamefont
  {Hoang}, \citenamefont {Han}, \citenamefont {Ma},\ and\ \citenamefont
  {Li}}]{Ahn2018}%
  \BibitemOpen
  \bibfield  {author} {\bibinfo {author} {\bibfnamefont {J.}~\bibnamefont
  {Ahn}}, \bibinfo {author} {\bibfnamefont {Z.}~\bibnamefont {Xu}}, \bibinfo
  {author} {\bibfnamefont {J.}~\bibnamefont {Bang}}, \bibinfo {author}
  {\bibfnamefont {Y.-H.}\ \bibnamefont {Deng}}, \bibinfo {author}
  {\bibfnamefont {T.~M.}\ \bibnamefont {Hoang}}, \bibinfo {author}
  {\bibfnamefont {Q.}~\bibnamefont {Han}}, \bibinfo {author} {\bibfnamefont
  {R.-M.}\ \bibnamefont {Ma}},\ and\ \bibinfo {author} {\bibfnamefont
  {T.}~\bibnamefont {Li}},\ }\href
  {https://doi.org/10.1103/PhysRevLett.121.033603} {\bibfield  {journal}
  {\bibinfo  {journal} {Phys. Rev. Lett.}\ }\textbf {\bibinfo {volume} {121}},\
  \bibinfo {pages} {033603} (\bibinfo {year} {2018})}\BibitemShut {NoStop}%
\bibitem [{\citenamefont {Kim}\ \emph {et~al.}(2016)\citenamefont {Kim},
  \citenamefont {Hauer}, \citenamefont {Doolin}, \citenamefont {Souris},\ and\
  \citenamefont {Davis}}]{Kim2016}%
  \BibitemOpen
  \bibfield  {author} {\bibinfo {author} {\bibfnamefont {P.~H.}\ \bibnamefont
  {Kim}}, \bibinfo {author} {\bibfnamefont {B.~D.}\ \bibnamefont {Hauer}},
  \bibinfo {author} {\bibfnamefont {C.}~\bibnamefont {Doolin}}, \bibinfo
  {author} {\bibfnamefont {F.}~\bibnamefont {Souris}},\ and\ \bibinfo {author}
  {\bibfnamefont {J.~P.}\ \bibnamefont {Davis}},\ }\href
  {https://doi.org/10.1038/ncomms13165} {\bibfield  {journal} {\bibinfo
  {journal} {Nat. Commun.}\ }\textbf {\bibinfo {volume} {7}},\ \bibinfo {pages}
  {13165} (\bibinfo {year} {2016})}\BibitemShut {NoStop}%
\bibitem [{\citenamefont {Haiberger}\ \emph {et~al.}(2007)\citenamefont
  {Haiberger}, \citenamefont {Weingran},\ and\ \citenamefont
  {Schiller}}]{Haiberger2007}%
  \BibitemOpen
  \bibfield  {author} {\bibinfo {author} {\bibfnamefont {L.}~\bibnamefont
  {Haiberger}}, \bibinfo {author} {\bibfnamefont {M.}~\bibnamefont
  {Weingran}},\ and\ \bibinfo {author} {\bibfnamefont {S.}~\bibnamefont
  {Schiller}},\ }\href {https://doi.org/10.1063/1.2437133} {\bibfield
  {journal} {\bibinfo  {journal} {Rev. Sci. Instrum.}\ }\textbf {\bibinfo
  {volume} {78}},\ \bibinfo {pages} {025101} (\bibinfo {year}
  {2007})}\BibitemShut {NoStop}%
\bibitem [{\citenamefont {Mueller}\ \emph
  {et~al.}(2008{\natexlab{a}})\citenamefont {Mueller}, \citenamefont {Heugel},\
  and\ \citenamefont {Wang}}]{Mueller2008b}%
  \BibitemOpen
  \bibfield  {author} {\bibinfo {author} {\bibfnamefont {F.}~\bibnamefont
  {Mueller}}, \bibinfo {author} {\bibfnamefont {S.}~\bibnamefont {Heugel}},\
  and\ \bibinfo {author} {\bibfnamefont {L.~J.}\ \bibnamefont {Wang}},\ }\href
  {https://doi.org/10.1364/OL.33.000539} {\bibfield  {journal} {\bibinfo
  {journal} {Opt. Lett.}\ }\textbf {\bibinfo {volume} {33}},\ \bibinfo {pages}
  {539} (\bibinfo {year} {2008}{\natexlab{a}})}\BibitemShut {NoStop}%
\bibitem [{\citenamefont {Cavalleri}\ \emph {et~al.}(2009)\citenamefont
  {Cavalleri}, \citenamefont {Ciani}, \citenamefont {Dolesi}, \citenamefont
  {Heptonstall}, \citenamefont {Hueller}, \citenamefont {Nicolodi},
  \citenamefont {Rowan}, \citenamefont {Tombolato}, \citenamefont {Vitale},
  \citenamefont {Wass},\ and\ \citenamefont {Weber}}]{Cavalleri2009}%
  \BibitemOpen
  \bibfield  {author} {\bibinfo {author} {\bibfnamefont {A.}~\bibnamefont
  {Cavalleri}}, \bibinfo {author} {\bibfnamefont {G.}~\bibnamefont {Ciani}},
  \bibinfo {author} {\bibfnamefont {R.}~\bibnamefont {Dolesi}}, \bibinfo
  {author} {\bibfnamefont {A.}~\bibnamefont {Heptonstall}}, \bibinfo {author}
  {\bibfnamefont {M.}~\bibnamefont {Hueller}}, \bibinfo {author} {\bibfnamefont
  {D.}~\bibnamefont {Nicolodi}}, \bibinfo {author} {\bibfnamefont
  {S.}~\bibnamefont {Rowan}}, \bibinfo {author} {\bibfnamefont
  {D.}~\bibnamefont {Tombolato}}, \bibinfo {author} {\bibfnamefont
  {S.}~\bibnamefont {Vitale}}, \bibinfo {author} {\bibfnamefont {P.~J.}\
  \bibnamefont {Wass}},\ and\ \bibinfo {author} {\bibfnamefont {W.~J.}\
  \bibnamefont {Weber}},\ }\href
  {https://doi.org/10.1088/0264-9381/26/9/094017} {\bibfield  {journal}
  {\bibinfo  {journal} {Classical and Quantum Gravity}\ }\textbf {\bibinfo
  {volume} {26}},\ \bibinfo {pages} {094017} (\bibinfo {year}
  {2009})}\BibitemShut {NoStop}%
\bibitem [{\citenamefont {Mueller}\ \emph
  {et~al.}(2008{\natexlab{b}})\citenamefont {Mueller}, \citenamefont {Heugel},\
  and\ \citenamefont {Wang}}]{Mueller2008a}%
  \BibitemOpen
  \bibfield  {author} {\bibinfo {author} {\bibfnamefont {F.}~\bibnamefont
  {Mueller}}, \bibinfo {author} {\bibfnamefont {S.}~\bibnamefont {Heugel}},\
  and\ \bibinfo {author} {\bibfnamefont {L.~J.}\ \bibnamefont {Wang}},\ }\href
  {https://doi.org/10.1103/PhysRevA.77.031802} {\bibfield  {journal} {\bibinfo
  {journal} {Phys. Rev. A}\ }\textbf {\bibinfo {volume} {77}},\ \bibinfo
  {pages} {031802} (\bibinfo {year} {2008}{\natexlab{b}})}\BibitemShut
  {NoStop}%
\bibitem [{\citenamefont {McManus}\ \emph {et~al.}(2017)\citenamefont
  {McManus}, \citenamefont {Forsyth}, \citenamefont {Yap}, \citenamefont
  {Ward}, \citenamefont {Shaddock}, \citenamefont {McClelland},\ and\
  \citenamefont {Slagmolen}}]{McManus2017}%
  \BibitemOpen
  \bibfield  {author} {\bibinfo {author} {\bibfnamefont {D.~J.}\ \bibnamefont
  {McManus}}, \bibinfo {author} {\bibfnamefont {P.~W.~F.}\ \bibnamefont
  {Forsyth}}, \bibinfo {author} {\bibfnamefont {M.~J.}\ \bibnamefont {Yap}},
  \bibinfo {author} {\bibfnamefont {R.~L.}\ \bibnamefont {Ward}}, \bibinfo
  {author} {\bibfnamefont {D.~A.}\ \bibnamefont {Shaddock}}, \bibinfo {author}
  {\bibfnamefont {D.~E.}\ \bibnamefont {McClelland}},\ and\ \bibinfo {author}
  {\bibfnamefont {B.~J.~J.}\ \bibnamefont {Slagmolen}},\ }\href
  {https://doi.org/10.1088/1361-6382/aa7103} {\bibfield  {journal} {\bibinfo
  {journal} {Classical and Quantum Gravity}\ }\textbf {\bibinfo {volume}
  {34}},\ \bibinfo {pages} {135002} (\bibinfo {year} {2017})}\BibitemShut
  {NoStop}%
\bibitem [{\citenamefont {Tan}\ \emph {et~al.}(2015)\citenamefont {Tan},
  \citenamefont {Hu},\ and\ \citenamefont {Shao}}]{Tan2015}%
  \BibitemOpen
  \bibfield  {author} {\bibinfo {author} {\bibfnamefont {Y.-J.}\ \bibnamefont
  {Tan}}, \bibinfo {author} {\bibfnamefont {Z.-K.}\ \bibnamefont {Hu}},\ and\
  \bibinfo {author} {\bibfnamefont {C.-G.}\ \bibnamefont {Shao}},\ }\href
  {https://doi.org/10.1103/PhysRevA.92.032131} {\bibfield  {journal} {\bibinfo
  {journal} {Phys. Rev. A}\ }\textbf {\bibinfo {volume} {92}},\ \bibinfo
  {pages} {032131} (\bibinfo {year} {2015})}\BibitemShut {NoStop}%
\bibitem [{\citenamefont {Li}\ \emph {et~al.}(2018)\citenamefont {Li},
  \citenamefont {Xue}, \citenamefont {Liu}, \citenamefont {Wu}, \citenamefont
  {Yang}, \citenamefont {Shao}, \citenamefont {Quan}, \citenamefont {Tan},
  \citenamefont {Tu}, \citenamefont {Liu}, \citenamefont {Xu}, \citenamefont
  {Liu}, \citenamefont {Wang}, \citenamefont {Hu}, \citenamefont {Zhou},
  \citenamefont {Luo}, \citenamefont {Wu}, \citenamefont {Milyukov},\ and\
  \citenamefont {Luo}}]{Li2018}%
  \BibitemOpen
  \bibfield  {author} {\bibinfo {author} {\bibfnamefont {Q.}~\bibnamefont
  {Li}}, \bibinfo {author} {\bibfnamefont {C.}~\bibnamefont {Xue}}, \bibinfo
  {author} {\bibfnamefont {J.-P.}\ \bibnamefont {Liu}}, \bibinfo {author}
  {\bibfnamefont {J.-F.}\ \bibnamefont {Wu}}, \bibinfo {author} {\bibfnamefont
  {S.-Q.}\ \bibnamefont {Yang}}, \bibinfo {author} {\bibfnamefont {C.-G.}\
  \bibnamefont {Shao}}, \bibinfo {author} {\bibfnamefont {L.-D.}\ \bibnamefont
  {Quan}}, \bibinfo {author} {\bibfnamefont {W.-H.}\ \bibnamefont {Tan}},
  \bibinfo {author} {\bibfnamefont {L.-C.}\ \bibnamefont {Tu}}, \bibinfo
  {author} {\bibfnamefont {Q.}~\bibnamefont {Liu}}, \bibinfo {author}
  {\bibfnamefont {H.}~\bibnamefont {Xu}}, \bibinfo {author} {\bibfnamefont
  {L.-X.}\ \bibnamefont {Liu}}, \bibinfo {author} {\bibfnamefont {Q.-L.}\
  \bibnamefont {Wang}}, \bibinfo {author} {\bibfnamefont {Z.-K.}\ \bibnamefont
  {Hu}}, \bibinfo {author} {\bibfnamefont {Z.-B.}\ \bibnamefont {Zhou}},
  \bibinfo {author} {\bibfnamefont {P.-S.}\ \bibnamefont {Luo}}, \bibinfo
  {author} {\bibfnamefont {S.-C.}\ \bibnamefont {Wu}}, \bibinfo {author}
  {\bibfnamefont {V.}~\bibnamefont {Milyukov}},\ and\ \bibinfo {author}
  {\bibfnamefont {J.}~\bibnamefont {Luo}},\ }\href
  {https://doi.org/10.1038/s41586-018-0431-5} {\bibfield  {journal} {\bibinfo
  {journal} {Nature (London)}\ }\textbf {\bibinfo {volume} {560}},\ \bibinfo
  {pages} {582} (\bibinfo {year} {2018})}\BibitemShut {NoStop}%
\bibitem [{\citenamefont {Matsumoto}\ \emph {et~al.}(2014)\citenamefont
  {Matsumoto}, \citenamefont {Michimura}, \citenamefont {Aso},\ and\
  \citenamefont {Tsubono}}]{Matsumoto2014}%
  \BibitemOpen
  \bibfield  {author} {\bibinfo {author} {\bibfnamefont {N.}~\bibnamefont
  {Matsumoto}}, \bibinfo {author} {\bibfnamefont {Y.}~\bibnamefont
  {Michimura}}, \bibinfo {author} {\bibfnamefont {Y.}~\bibnamefont {Aso}},\
  and\ \bibinfo {author} {\bibfnamefont {K.}~\bibnamefont {Tsubono}},\ }\href
  {https://doi.org/10.1364/OE.22.012915} {\bibfield  {journal} {\bibinfo
  {journal} {Opt. Express}\ }\textbf {\bibinfo {volume} {22}},\ \bibinfo
  {pages} {12915} (\bibinfo {year} {2014})}\BibitemShut {NoStop}%
\bibitem [{the()}]{thermalnoisecal}%
  \BibitemOpen
  \href@noop {} {\bibinfo  {journal} {The thermal noise contribution is
  estimated from the calculated effective mass of 26\,mg, known resonant
  frequency of 5.8\,kHz, and the observed $Q$ of $10^4$ (limited by coating and
  substrate surface losses)}\ }\BibitemShut {NoStop}%
\bibitem [{\citenamefont {Matsumoto}\ \emph {et~al.}(2019)\citenamefont
  {Matsumoto}, \citenamefont {Cata\~no Lopez}, \citenamefont {Sugawara},
  \citenamefont {Suzuki}, \citenamefont {Abe}, \citenamefont {Komori},
  \citenamefont {Michimura}, \citenamefont {Aso},\ and\ \citenamefont
  {Edamatsu}}]{Matsumoto2019}%
  \BibitemOpen
\bibfield  {journal} {  }\bibfield  {author} {\bibinfo {author} {\bibfnamefont
  {N.}~\bibnamefont {Matsumoto}}, \bibinfo {author} {\bibfnamefont {S.~B.}\
  \bibnamefont {Cata\~no Lopez}}, \bibinfo {author} {\bibfnamefont
  {M.}~\bibnamefont {Sugawara}}, \bibinfo {author} {\bibfnamefont
  {S.}~\bibnamefont {Suzuki}}, \bibinfo {author} {\bibfnamefont
  {N.}~\bibnamefont {Abe}}, \bibinfo {author} {\bibfnamefont {K.}~\bibnamefont
  {Komori}}, \bibinfo {author} {\bibfnamefont {Y.}~\bibnamefont {Michimura}},
  \bibinfo {author} {\bibfnamefont {Y.}~\bibnamefont {Aso}},\ and\ \bibinfo
  {author} {\bibfnamefont {K.}~\bibnamefont {Edamatsu}},\ }\href
  {https://doi.org/10.1103/PhysRevLett.122.071101} {\bibfield  {journal}
  {\bibinfo  {journal} {Phys. Rev. Lett.}\ }\textbf {\bibinfo {volume} {122}},\
  \bibinfo {pages} {071101} (\bibinfo {year} {2019})}\BibitemShut {NoStop}%
\bibitem [{\citenamefont {Neben}\ \emph {et~al.}(2012)\citenamefont {Neben},
  \citenamefont {Bodiya}, \citenamefont {Wipf}, \citenamefont {Oelker},
  \citenamefont {Corbitt},\ and\ \citenamefont {Mavalvala}}]{Neben2012}%
  \BibitemOpen
  \bibfield  {author} {\bibinfo {author} {\bibfnamefont {A.~R.}\ \bibnamefont
  {Neben}}, \bibinfo {author} {\bibfnamefont {T.~P.}\ \bibnamefont {Bodiya}},
  \bibinfo {author} {\bibfnamefont {C.}~\bibnamefont {Wipf}}, \bibinfo {author}
  {\bibfnamefont {E.}~\bibnamefont {Oelker}}, \bibinfo {author} {\bibfnamefont
  {T.}~\bibnamefont {Corbitt}},\ and\ \bibinfo {author} {\bibfnamefont
  {N.}~\bibnamefont {Mavalvala}},\ }\href
  {https://doi.org/10.1088/1367-2630/14/11/115008} {\bibfield  {journal}
  {\bibinfo  {journal} {New J. Phys.}\ }\textbf {\bibinfo {volume} {14}},\
  \bibinfo {pages} {115008} (\bibinfo {year} {2012})}\BibitemShut {NoStop}%
\bibitem [{\citenamefont {Matsumoto}\ \emph {et~al.}(2015)\citenamefont
  {Matsumoto}, \citenamefont {Komori}, \citenamefont {Michimura}, \citenamefont
  {Hayase}, \citenamefont {Aso},\ and\ \citenamefont
  {Tsubono}}]{Matsumoto2015}%
  \BibitemOpen
  \bibfield  {author} {\bibinfo {author} {\bibfnamefont {N.}~\bibnamefont
  {Matsumoto}}, \bibinfo {author} {\bibfnamefont {K.}~\bibnamefont {Komori}},
  \bibinfo {author} {\bibfnamefont {Y.}~\bibnamefont {Michimura}}, \bibinfo
  {author} {\bibfnamefont {G.}~\bibnamefont {Hayase}}, \bibinfo {author}
  {\bibfnamefont {Y.}~\bibnamefont {Aso}},\ and\ \bibinfo {author}
  {\bibfnamefont {K.}~\bibnamefont {Tsubono}},\ }\href
  {https://doi.org/10.1103/PhysRevA.92.033825} {\bibfield  {journal} {\bibinfo
  {journal} {Phys. Rev. A}\ }\textbf {\bibinfo {volume} {92}},\ \bibinfo
  {pages} {033825} (\bibinfo {year} {2015})}\BibitemShut {NoStop}%
\bibitem [{\citenamefont {Nguyen}\ \emph {et~al.}(2015)\citenamefont {Nguyen},
  \citenamefont {Mow-Lowry}, \citenamefont {Slagmolen}, \citenamefont {Miller},
  \citenamefont {Mullavey}, \citenamefont {Go\ss{}ler}, \citenamefont {Altin},
  \citenamefont {Shaddock},\ and\ \citenamefont {McClelland}}]{Nguyen2015}%
  \BibitemOpen
  \bibfield  {author} {\bibinfo {author} {\bibfnamefont {T.~T.-H.}\
  \bibnamefont {Nguyen}}, \bibinfo {author} {\bibfnamefont {C.~M.}\
  \bibnamefont {Mow-Lowry}}, \bibinfo {author} {\bibfnamefont {B.~J.~J.}\
  \bibnamefont {Slagmolen}}, \bibinfo {author} {\bibfnamefont {J.}~\bibnamefont
  {Miller}}, \bibinfo {author} {\bibfnamefont {A.~J.}\ \bibnamefont
  {Mullavey}}, \bibinfo {author} {\bibfnamefont {S.}~\bibnamefont
  {Go\ss{}ler}}, \bibinfo {author} {\bibfnamefont {P.~A.}\ \bibnamefont
  {Altin}}, \bibinfo {author} {\bibfnamefont {D.~A.}\ \bibnamefont
  {Shaddock}},\ and\ \bibinfo {author} {\bibfnamefont {D.~E.}\ \bibnamefont
  {McClelland}},\ }\href {https://doi.org/10.1103/PhysRevD.92.112004}
  {\bibfield  {journal} {\bibinfo  {journal} {Phys. Rev. D}\ }\textbf {\bibinfo
  {volume} {92}},\ \bibinfo {pages} {112004} (\bibinfo {year}
  {2015})}\BibitemShut {NoStop}%
\bibitem [{\citenamefont {Adler}(2007)}]{Adler2007}%
  \BibitemOpen
  \bibfield  {author} {\bibinfo {author} {\bibfnamefont {S.~L.}\ \bibnamefont
  {Adler}},\ }\href {https://doi.org/10.1088/1751-8113/40/12/s03} {\bibfield
  {journal} {\bibinfo  {journal} {Journal of Physics A: Mathematical and
  Theoretical}\ }\textbf {\bibinfo {volume} {40}},\ \bibinfo {pages} {2935}
  (\bibinfo {year} {2007})}\BibitemShut {NoStop}%
\bibitem [{\citenamefont {Vinante}\ \emph {et~al.}(2016)\citenamefont
  {Vinante}, \citenamefont {Bahrami}, \citenamefont {Bassi}, \citenamefont
  {Usenko}, \citenamefont {Wijts},\ and\ \citenamefont
  {Oosterkamp}}]{Vinante2016}%
  \BibitemOpen
  \bibfield  {author} {\bibinfo {author} {\bibfnamefont {A.}~\bibnamefont
  {Vinante}}, \bibinfo {author} {\bibfnamefont {M.}~\bibnamefont {Bahrami}},
  \bibinfo {author} {\bibfnamefont {A.}~\bibnamefont {Bassi}}, \bibinfo
  {author} {\bibfnamefont {O.}~\bibnamefont {Usenko}}, \bibinfo {author}
  {\bibfnamefont {G.}~\bibnamefont {Wijts}},\ and\ \bibinfo {author}
  {\bibfnamefont {T.~H.}\ \bibnamefont {Oosterkamp}},\ }\href
  {https://doi.org/10.1103/PhysRevLett.116.090402} {\bibfield  {journal}
  {\bibinfo  {journal} {Phys. Rev. Lett.}\ }\textbf {\bibinfo {volume} {116}},\
  \bibinfo {pages} {090402} (\bibinfo {year} {2016})}\BibitemShut {NoStop}%
\bibitem [{\citenamefont {Vinante}\ \emph {et~al.}(2017)\citenamefont
  {Vinante}, \citenamefont {Mezzena}, \citenamefont {Falferi}, \citenamefont
  {Carlesso},\ and\ \citenamefont {Bassi}}]{Vinante2017}%
  \BibitemOpen
  \bibfield  {author} {\bibinfo {author} {\bibfnamefont {A.}~\bibnamefont
  {Vinante}}, \bibinfo {author} {\bibfnamefont {R.}~\bibnamefont {Mezzena}},
  \bibinfo {author} {\bibfnamefont {P.}~\bibnamefont {Falferi}}, \bibinfo
  {author} {\bibfnamefont {M.}~\bibnamefont {Carlesso}},\ and\ \bibinfo
  {author} {\bibfnamefont {A.}~\bibnamefont {Bassi}},\ }\href
  {https://doi.org/10.1103/PhysRevLett.119.110401} {\bibfield  {journal}
  {\bibinfo  {journal} {Phys. Rev. Lett.}\ }\textbf {\bibinfo {volume} {119}},\
  \bibinfo {pages} {110401} (\bibinfo {year} {2017})}\BibitemShut {NoStop}%
\bibitem [{\citenamefont {Carlesso}\ \emph {et~al.}(2016)\citenamefont
  {Carlesso}, \citenamefont {Bassi}, \citenamefont {Falferi},\ and\
  \citenamefont {Vinante}}]{Carlesso2016}%
  \BibitemOpen
  \bibfield  {author} {\bibinfo {author} {\bibfnamefont {M.}~\bibnamefont
  {Carlesso}}, \bibinfo {author} {\bibfnamefont {A.}~\bibnamefont {Bassi}},
  \bibinfo {author} {\bibfnamefont {P.}~\bibnamefont {Falferi}},\ and\ \bibinfo
  {author} {\bibfnamefont {A.}~\bibnamefont {Vinante}},\ }\href
  {https://doi.org/10.1103/PhysRevD.94.124036} {\bibfield  {journal} {\bibinfo
  {journal} {Phys. Rev. D}\ }\textbf {\bibinfo {volume} {94}},\ \bibinfo
  {pages} {124036} (\bibinfo {year} {2016})}\BibitemShut {NoStop}%
\bibitem [{\citenamefont {Helou}\ \emph
  {et~al.}(2017{\natexlab{b}})\citenamefont {Helou}, \citenamefont {Slagmolen},
  \citenamefont {McClelland},\ and\ \citenamefont {Chen}}]{Helou2017b}%
  \BibitemOpen
  \bibfield  {author} {\bibinfo {author} {\bibfnamefont {B.}~\bibnamefont
  {Helou}}, \bibinfo {author} {\bibfnamefont {B.~J.~J.}\ \bibnamefont
  {Slagmolen}}, \bibinfo {author} {\bibfnamefont {D.~E.}\ \bibnamefont
  {McClelland}},\ and\ \bibinfo {author} {\bibfnamefont {Y.}~\bibnamefont
  {Chen}},\ }\href {https://doi.org/10.1103/PhysRevD.95.084054} {\bibfield
  {journal} {\bibinfo  {journal} {Phys. Rev. D}\ }\textbf {\bibinfo {volume}
  {95}},\ \bibinfo {pages} {084054} (\bibinfo {year}
  {2017}{\natexlab{b}})}\BibitemShut {NoStop}%
\end{thebibliography}%
\bibliographystyle{apsrev.bst}
\end{document}